\documentclass[onecolumn,aps,prd,superscriptaddress,preprintnumbers,nofootinbib,10pt]{revtex4-2}

\usepackage{amsfonts}
\usepackage[tbtags]{amsmath}
\usepackage{amssymb}
\usepackage{booktabs}
\usepackage{xcolor}
\usepackage{dsfont}
\usepackage[hidelinks]{hyperref}
\usepackage[utf8]{inputenc}
\usepackage{url}
\usepackage{orcidlink}

\allowdisplaybreaks

\hypersetup{
	colorlinks=true,
	citecolor=blue,
	citebordercolor=red,
	linktoc=all,
	linkcolor=blue,
	urlcolor=blue
}

\newcommand{\cA}{\cal A}
\newcommand{\cB}{\cal B}
\newcommand{\cC}{\cal C}
\newcommand{\cM}{\cal M}

\renewcommand{\Im}{\mathrm{Im}}

\begin{document}

	\title{Finite Weyl polynomials and the approach to Tsirelson’s bound in relativistic scalar quantum field theory}

	\author{J. G. A. Carib{\'e}~\orcidlink{https://orcid.org/0000-0001-7418-5149}}\email{joaogcaribe@gmail.com} \affiliation{UERJ $-$ Universidade do Estado do Rio de Janeiro, Instituto de Física $-$ Departamento de Física Teórica $-$ Rua São Francisco Xavier 524, 20550-013, Maracanã, Rio de Janeiro, Brazil}

	\author{M. S. Guimaraes~\orcidlink{0000-0003-0057-148X}}\email{msguimaraes@uerj.br} \affiliation{UERJ $-$ Universidade do Estado do Rio de Janeiro, Instituto de Física $-$ Departamento de Física Teórica $-$ Rua São Francisco Xavier 524, 20550-013, Maracanã, Rio de Janeiro, Brazil}

	\author{I. Roditi~\orcidlink{0000-0003-2363-5626}}\email{roditi@cbpf.br} \affiliation{CBPF $-$ Centro Brasileiro de Pesquisas Físicas, Rua Dr. Xavier Sigaud 150, 22290-180, Rio de Janeiro, Brazil}

	\author{S. P. Sorella~\orcidlink{0000-0002-6051-6960}}\email{silvio.sorella@fis.uerj.br} \affiliation{UERJ $-$ Universidade do Estado do Rio de Janeiro, Instituto de Física $-$ Departamento de Física Teórica $-$ Rua São Francisco Xavier 524, 20550-013, Maracanã, Rio de Janeiro, Brazil}

	\begin{abstract}

	We construct four bounded Hermitian operators, each one a finite polynomial in the unitary Weyl operators, for the Bell-CHSH inequality in a free massive scalar field in $1+1$ dimensions. The operators are localized in complementary wedges. Odd Weyl harmonics provide two exactly anticommuting axis observables in each wedge, while normalizable packets with compact support in the spectrum of the boost generator give exact modular inner products at nonvanishing bandwidth. The resulting Bell-CHSH correlator is a finite double sum. An operator with six Weyl terms per axis already gives $2.14885$. Using normalized Fejér approximants of $\operatorname{sgn}(\cos(x))$, we show that the supremum over the finite-polynomial family equals $2\sqrt{2}$, although no finite member attains it; a degree-$511$ example gives $2.80027$. We also point out that, in a centered quasifree state, a Bell-CHSH test whose four final settings are bounded functions of individual quadratures admits one common Gaussian representation and therefore stays below $2$. The finite-Weyl construction avoids this restriction because Bob's final settings mix two noncommuting axis observables. 
	\end{abstract}

	\maketitle

\section{Introduction}\label{intro}

The Summers-Werner results show that the vacuum Bell-correlation supremum can equal $2\sqrt{2}$ for suitable space-like separated regions in free Bose and Fermi theories \cite{Summers:1985wq,Summers:1987fn,Summers:1987squ,Summers:1987ze}. Their analysis establishes the existence of bounded local operators with maximal Bell correlations. Though, it does not provide simple bosonic observables or a finite set of Weyl coefficients. The aim of the present work is to supply such data for the massive scalar field in $1+1$ dimensions.

The problem is already visible at the level of the local field. The smeared scalar field $\varphi(f)$ is unbounded, whereas the most immediate bounded replacements, such as $\cos\varphi(f)$, $\tanh\varphi(f)$ and the spectral projections of $\varphi(f)$, are functions of one quadrature, {\it i.e.}, a smeared field for a fixed real one-particle direction $f$. In a centered quasifree state, a Bell-CHSH test in which each of the four final settings has this form admits one common positive Gaussian representation. The Hadamard cross-correlations are fully retained as off-diagonal covariances, but the Bell-CHSH value cannot exceed $2$. This observation will be recalled briefly in Sect.~\ref{gauss}.

The full local Weyl algebra is not restricted in this way. Weyl operators associated with different local directions retain the symplectic phase of eq.~\eqref{weylalg}, so a finite Hermitian sum over two noncommuting directions is not a bounded function of a single quadrature and admits no common Gaussian representation. The distinction matters: a finite Weyl sum along one direction is again a function of that quadrature and stays inside the class just described. Recent constructions have combined Weyl operators and modular localization in the study of Bell correlations in QFT \cite{DeFabritiis:2023tkh,Caribe:2026mod}. Here we shall use the symplectic phase to build, in each wedge, two axis observables that anticommute exactly. Assigning the complementary wedges to Alice and Bob, Alice's two settings are her axis observables themselves, while Bob's are their normalized sum and difference. Anticommutation is what keeps the latter bounded by one, and each of them superposes bounded functions of two conjugate quadratures, so Bob's settings fall outside the one-quadrature class recalled above.

The construction has three ingredients. First, a bounded trigonometric polynomial containing only odd harmonics gives a finite Hermitian Weyl polynomial. For symplectically normalized directions, two such axis observables anticommute exactly. Second, modular localization provides two pairs of one-particle vectors in complementary wedges. We use a normalizable packet with compact support in the spectrum of the boost generator, so that all the relevant inner products are exact at nonzero bandwidth. Third, normalized Fejér approximants of $\operatorname{sgn}(\cos( x))$ supply a finite-polynomial sequence whose Bell-CHSH values approach $2\sqrt{2}$.

More precisely, for every $\varepsilon>0$ we obtain a finite Bell-CHSH quadruple satisfying
\begin{equation}
2\sqrt{2}-\varepsilon
<
\langle0|{\cal C}|0\rangle
<
2\sqrt{2}\;.
\label{introeps}
\end{equation}
No finite member attains the supremum. An economical polynomial with six Weyl operators per axis gives $2.14885$ at an exact finite-band point, while a degree-$511$ example gives $2.80027$. The construction remains valid after spectral smoothing.  

In the narrow-band limit the covariance data reduce to those of a two-mode squeezed vacuum. This provides a direct check of the formulas and connects the construction with the parity-pseudospin approach of Ref.~\cite{Chen:2002cv}. The distinction is that the present settings are finite Weyl polynomials and are localized in complementary wedges at every nonvanishing bandwidth. A different bosonic construction based on chiral vertex operators was presented in Ref.~\cite{Caribe:2026vtx}.

Sect.~\ref{sec:framework} fixes the Bell-CHSH and CCR conventions and recalls the Gaussian restriction for four final one-quadrature settings. Sect.~\ref{sec:algebraic} gives the finite Weyl polynomials. The modular vectors are constructed in Sect.~\ref{sec:modular}, and Sect.~\ref{sec:construction} contains the correlator, the explicit violations and the approach to Tsirelson's bound. Technical steps are collected in the appendices.

The appendices form an essential part of the analysis. Appendix~\ref{appGram} derives the finite-band modular inner products in the boost-spectral representation and establishes the formulas used throughout the construction. Appendix~\ref{appZero} analyzes the modular zero-spectrum limit, proving the limiting behavior of the exponent combinations without imposing an additional narrow-band assumption. Appendix~\ref{appPoly} supplies the bounds on the trigonometric polynomials, including the Fej\'er normalization and the six-Weyl example. Appendix~\ref{app:p511} gives the explicit degree-$511$ Fej\'er polynomial, its finite-band Bell-CHSH value, and the numerically stable form used for its evaluation. Finally, Appendix~\ref{app:abel-poisson} discusses Abel--Poisson approximants as an alternative regularization, showing that the odd-harmonic Weyl-polynomial construction is not tied to a unique summability prescription.

\section{The Bell-CHSH inequality for commuting local algebras}\label{sec:framework}

\subsection{Conventions}\label{conv}

Let ${\cal H}_1$ be the one-particle Hilbert space of a free real scalar field, the inner product $\langle h | k \rangle$ being linear in the second entry. The symplectic form is provided by the smeared Pauli-Jordan distribution
\begin{equation}
\Delta_{PJ}(h,k) = 2\, \Im \langle h | k \rangle \;, \qquad \left[ \varphi(h), \varphi(k) \right] = i \Delta_{PJ}(h,k) \;. \label{PJdef}
\end{equation}
The unitary Weyl operators
\begin{equation}
W(h) = e^{i \varphi(h)} \;, \qquad W^\dagger(h) = W(-h) \;, \label{Wdef}
\end{equation}
obey
\begin{eqnarray}
W(h) W(k) & = & e^{-\frac{i}{2} \Delta_{PJ}(h,k)} \; W(h+k) \;, \nonumber \\
W(h) W(k) & = & e^{-i \Delta_{PJ}(h,k)} \; W(k) W(h) \;, \label{weylalg}
\end{eqnarray}
while, in the Fock vacuum,
\begin{equation}
\langle 0|\; W(h) \;|0\rangle = e^{-\frac{1}{2} ||h||^2} \;, \qquad ||h||^2 = \langle h|h\rangle \;. \label{vacW}
\end{equation}
All numerical values quoted below refer to these conventions. The modular operator of the one-particle space will be denoted by $\delta$, the subscript $PJ$ distinguishing the Pauli-Jordan distribution from the modular operator.

\subsection{The operator identity and the contraction bound}\label{opid}

Let ${\cA}$ and ${\cB}$ be commuting von Neumann algebras localized in space-like separated regions, and let $(A,A') \in {\cA}$, $(B,B') \in {\cB}$ be Hermitian operators with $||A||,||A'||,||B||,||B'|| \le 1$. The Bell-CHSH operator is \cite{Bell:1964kc,Clauser:1969ny,Cirelson:1980ry}
\begin{equation}
{\cC} = (A+A')B + (A-A')B' \;. \label{CHSHop}
\end{equation}
Whenever the four settings admit one common positive probability representation with responses in $[-1,1]$,
\begin{equation}
\omega(AB) = \int_\Lambda a(\lambda) b(\lambda) \, d\nu(\lambda) \;, \qquad |a|,|b| \le 1 \;, \label{lhv}
\end{equation}
one has $|\omega({\cC})| \le 2$. For binary outcomes, the relation between a common joint probability distribution, factorizable response models and the Bell inequalities was made precise by Fine \cite{Fine:1982zz}.

For dichotomic settings, $A^2=A'^2=B^2=B'^2=\mathds{1}$, expanding \eqref{CHSHop} gives the well-known identity
\begin{equation}
{\cC}^2 = 4\,\mathds{1} - [A,A'][B,B'] \;, \label{tsid}
\end{equation}
whence $||{\cC}|| \le 2\sqrt{2}$, since the norm of each commutator is at most $2$. The operators constructed below are Hermitian contractions rather than involutions. In the GNS representation of the state $\omega$, with cyclic vector $\Omega$, the Cauchy-Schwarz inequality in ${\cal H} \oplus {\cal H}$ gives
\begin{equation}
|\omega({\cC})| \le \left( ||A\Omega||^2 + ||A'\Omega||^2 \right)^{1/2} \left( ||(B+B')\Omega||^2 + ||(B-B')\Omega||^2 \right)^{1/2} \;.
\end{equation}
The parallelogram identity then gives
\begin{equation}
|\omega({\cC})| \le \sqrt{\omega(A^2+A'^2)} \; \sqrt{2\, \omega(B^2+B'^2)} \le 2 \sqrt{2} \;, \label{contr}
\end{equation}
which is the form of Tsirelson's bound employed below. Equation \eqref{tsid} also makes clear that space-like locality, $[{\cA},{\cB}]=0$, is compatible with values above $2$. Each local pair must be noncommutative, although local noncommutativity alone is not sufficient: the state must correlate the two pairs in an appropriate way.

\subsection{One-quadrature operators and the common Gaussian measure}\label{gauss}

Let $\omega$ be a regular centered quasifree state of a CCR algebra, with real symmetric covariance $\mu$ defined by
\begin{equation}
\omega(W(h)) = e^{-\frac{1}{2} \mu(h,h)} \;. \label{qfree}
\end{equation}
For the Fock vacuum, $\mu$ is the Hadamard covariance. Consider a Bell-CHSH test whose four final settings are bounded Borel functions of individual smeared fields,
\begin{equation}
A(f) = F(\varphi(f)) \;, \qquad A(f')=F'(\varphi(f')) \;, \qquad B(g)=G(\varphi(g)) \;, \qquad B(g')=G'(\varphi(g')) \;, \label{oneq}
\end{equation}
with $F,F',G,G' : {\mathbb R} \to [-1,1]$, and let the supports be space-like, so that
\begin{equation}
\Delta_{PJ}(f,g)=\Delta_{PJ}(f,g')=\Delta_{PJ}(f',g)=\Delta_{PJ}(f',g')=0 \;. \label{spl}
\end{equation}
Setting $(h_1,h_2,h_3,h_4)=(f,f',g,g')$ and
\begin{equation}
\Gamma_{\alpha \beta} = \mu(h_\alpha,h_\beta) \;, \qquad \alpha,\beta=1,\dots,4 \;, \label{Gam}
\end{equation}
the matrix $\Gamma$ is positive semidefinite and therefore defines a centered Gaussian measure $\gamma_\Gamma$ on ${\mathbb R}^4$. For every $u=(u_1,\dots,u_4) \in {\mathbb R}^4$,
\begin{equation}
\int_{{\mathbb R}^4} e^{i u \cdot z} \, d\gamma_\Gamma(z) = \exp\left[ -\frac{1}{2} \mu\left( \sum_\alpha u_\alpha h_\alpha , \sum_\beta u_\beta h_\beta \right) \right] \;. \label{chf}
\end{equation}
For the context $(f,g)$, condition \eqref{spl} removes the Weyl phase and gives
\begin{equation}
\omega\left( e^{it \varphi(f)} e^{is\varphi(g)} \right) = \exp\left[ -\frac{1}{2}\left( t^2 \mu(f,f) + s^2 \mu(g,g) + 2ts\, \mu(f,g)\right) \right] \;, \label{ctx}
\end{equation}
which is the characteristic function of the $(z_1,z_3)$-marginal of $\gamma_\Gamma$. The uniqueness of characteristic functions and the bounded Borel functional calculus then give
\begin{equation}
\omega\left( A(f)B(g) \right) = \int_{{\mathbb R}^4} F(z_1) G(z_3) \, d\gamma_\Gamma(z) \;. \label{reprc}
\end{equation}
The remaining three contexts are obtained from the corresponding marginals of the same measure. Thus \eqref{reprc} has the form \eqref{lhv}, with responses $F(z_1)$, $F'(z_2)$, $G(z_3)$ and $G'(z_4)$, and
\begin{equation}
\left| \omega({\cC}) \right| \le 2 \;. \label{nogo}
\end{equation}
The Hadamard cross-correlations between the two regions have not been discarded: they are the off-diagonal entries of $\Gamma$. What gives the classical bound is the existence of one context-independent positive measure, not a factorization of the vacuum state. Notice also that the argument concerns the four selected cross-region contexts; it does not assign a joint quantum measurement to the generally noncommuting operators $A(f)$ and $A(f')$.

This restriction applies only when all four final settings have the form \eqref{oneq}. In the construction below, the two axis observables in each wedge remain functions of individual quadratures, but Bob's final settings are sums and differences of two noncommuting axes \footnote{ Notice that a setting means a bounded element of the local algebra, and the Bell-CHSH value is a vacuum expectation of such elements. Whether a given local observable can be measured by a physically realizable apparatus is a separate question and goes beyond our scope. This kind of question is addressed by the probe-based framework of Fewster and Verch \cite{Fewster:2018pey}, in which the causal pathologies of naive local state updates are absent \cite{Bostelmann:2020unl}. Since every local observable admits asymptotic measurement schemes in that framework \cite{Fewster:2022ori}, the finite Weyl polynomials constructed below are unproblematic in this respect.}. They are therefore outside the class \eqref{oneq}.

Writing a bounded function formally as
\begin{equation}
F(\varphi(f))
=
\int_{\mathbb R}\widehat F(k)W(kf)\,dk
\label{fourierrep}
\end{equation}
should not be confused with a probabilistic decomposition of the
observable. In general, the Fourier coefficient $\widehat F(k)$ is
neither positive nor even an ordinary function, and the corresponding
representation depends on the particular observable $F$. Thus,
Eq.~(18) does not provide the single, context-independent positive
measure required in Eq.~(7). In particular, when such a common representation exists
for the four one-quadrature settings considered above, its classical
random variables are the jointly Gaussian field amplitudes
$z_1,\ldots,z_4$ distributed according to $\gamma_\Gamma$.

\section{Finite Weyl polynomials and exact anticommutation}\label{sec:algebraic}

The local Weyl algebra is noncommutative: whenever $\Delta_{PJ}(h,k) \neq 0$, the phase appearing in the second of eqs.~\eqref{weylalg} survives, and a finite sum
\begin{equation}
A = \sum_r c_r W(h_r) \nonumber
\end{equation}
is not a probabilistic mixture. Let $M \subset \{1,3,5,\dots\}$ be a finite set of odd integers and let
\begin{equation}
p(x) = \sum_{m \in M} c_m \cos(mx) \;, \qquad c_m \in {\mathbb R} \;. \label{pdef}
\end{equation}
For a one-particle vector $h$ we define
\begin{equation}
S_p(h) = p\left( \alpha \varphi(h) \right) \;, \qquad \alpha = \sqrt{\frac{\pi}{2}} \;. \label{Sdef}
\end{equation}
Suppose that $||p||_\infty \le 1$ and that the two directions $(h,k)$ are symplectically normalized, {\it i.e.}
\begin{equation}
\Delta_{PJ}(h,k) = 2 \;, \qquad \Im \langle h|k\rangle = 1 \;. \label{norml}
\end{equation}
Then $S_p(h)$ and $S_p(k)$ are bounded Hermitian operators which anticommute exactly,
\begin{equation}
\{ S_p(h), S_p(k) \} = 0 \;. \label{anticomm}
\end{equation}
From the spectral theorem and $||p||_\infty \le 1$ it follows that $||S_p(h)||, ||S_p(k)|| \le 1$. Expanding each cosine into two Weyl operators,
\begin{equation}
S_p(h) = \frac{1}{2} \sum_{m \in M} c_m \left[ W(m \alpha h) + W(-m\alpha h) \right] \;, \label{SWeyl}
\end{equation}
so that both operators are finite Hermitian Weyl polynomials. Setting $U=W(\alpha h)$ and $V=W(\alpha k)$, eqs.~\eqref{weylalg} and \eqref{norml} give
\begin{equation}
UV = e^{-i \alpha^2 \Delta_{PJ}(h,k)} VU = e^{-i\pi} VU = -VU \;, \label{minus}
\end{equation}
the choice $\alpha^2=\pi/2$ being made for this purpose. For odd integers $m,n$, including the negative ones, one has $U^m V^n = - V^n U^m$. Since every harmonic entering \eqref{pdef} is odd, each term of $S_p(h)$ anticommutes with each term of $S_p(k)$, which proves \eqref{anticomm}.

This anticommutation allows Bob's two final settings to be assembled without spoiling boundedness. If $S$ and $T$ are anticommuting bounded Hermitian operators, then
\begin{equation}
B = \frac{S+T}{\sqrt{2}} \;, \qquad B' = \frac{S-T}{\sqrt{2}} \label{BBp}
\end{equation}
are again bounded Hermitian operators, since $\{S,T\}=0$ yields
\begin{equation}
B^2 = B'^2 = \frac{1}{2}\left( S^2+T^2\right) \le \mathds{1} \;. \nonumber
\end{equation}

\subsection{An economical polynomial: six Weyl operators per axis}\label{six}

The simplest polynomial which will be used below contains three odd harmonics only,
\begin{equation}
p_{\rm six}(x) = \frac{6}{5}\cos x - \frac{3}{10} \cos 3x + \frac{1}{10} \cos 5x \;. \label{psix}
\end{equation}
It obeys $||p_{\rm six}||_\infty = 1$; the short verification is given in Appendix~\ref{appPoly}. Accordingly,
\begin{eqnarray}
S_{\rm six}(h) & = & \frac{3}{5}\left[ W(\alpha h) + W(-\alpha h) \right] - \frac{3}{20}\left[ W(3\alpha h) + W(-3\alpha h)\right] \nonumber \\
& + & \frac{1}{20} \left[ W(5\alpha h) + W(-5\alpha h) \right] \label{Ssix}
\end{eqnarray}
is a bounded Hermitian operator containing {\it six} Weyl operators. Whenever $\Im\langle h|k\rangle=1$, the pair $\left( S_{\rm six}(h), S_{\rm six}(k)\right)$ anticommutes exactly.

\subsection{The normalized Fejér family}\label{fej}

The criterion \eqref{anticomm} admits a systematic family. Let
\begin{equation}
q(x) = \operatorname{sgn}(\cos (x) )\;, \qquad q(x) = \frac{4}{\pi} \sum_{k=0}^{\infty} \frac{(-1)^k}{2k+1} \cos\left( (2k+1)x\right) \;, \label{sgncos}
\end{equation}
the series being understood in $L^2([-\pi,\pi])$, and let $q_L = \sigma_{2L-1} q$ denote its Fejér mean of degree $2L-1$,
\begin{equation}
q_L(x) = \frac{4}{\pi} \sum_{k=0}^{L-1} \frac{(-1)^k}{2k+1}\left( 1 - \frac{2k+1}{2L}\right) \cos\left( (2k+1)x \right) \;. \label{fejer}
\end{equation}
Only odd harmonics appear, as required. Setting
\begin{equation}
M_L = q_L(0) \;, \qquad p_L(x) = \frac{q_L(x)}{M_L} = \sum_{k=0}^{L-1} c_{k,L} \cos\left((2k+1)x\right) \;, \label{pL}
\end{equation}
one shows, see Appendix~\ref{appPoly}, that
\begin{equation}
||q_L||_\infty = M_L \;, \qquad ||p_L||_\infty = 1 \;, \qquad \sum_{k=0}^{L-1} c_{k,L}^{\,2} \;\longrightarrow\; 2 \qquad (L \to \infty) \;. \label{pnorm}
\end{equation}
The value $2$ appearing as the limit in \eqref{pnorm} is not accidental: it is the largest value allowed by the boundedness of $p$. Indeed, for any polynomial of the form \eqref{pdef}, orthogonality of the cosines gives
\begin{equation}
\frac{1}{2\pi}\int_{-\pi}^{\pi} p(x)^2 dx = \frac{1}{2} \sum_{m \in M} c_m^2 \;,\nonumber
\end{equation}
so that $||p||_\infty \le 1$ implies at once the Parseval bound
\begin{equation}
\sum_{m \in M} c_m^2 \le 2 \;, \label{parseval}
\end{equation}
with equality if and only if $p(x)^2=1$ for almost every $x$. Since a finite trigonometric polynomial is continuous, equality would force $p$ to be constant, which is incompatible with the odd-harmonic structure. Hence \eqref{parseval} is strict for every finite $p$, and it becomes an equality only in the limit $p \to \operatorname{sgn}(\cos x)$.

We shall refer to $S_{p_L}(h)$ as an axis observable. It contains $2L$ Weyl operators, two for each of the $L$ odd harmonics. Every finite $L$ gives a bounded Hermitian operator, while the family as a whole carries the $L^2$ weight needed to approach Tsirelson's bound. For instance,
\begin{equation}
p_5(x) = \frac{567 \cos x - 147 \cos 3x + 63 \cos 5x - 27 \cos 7x + 7 \cos 9x}{463} \;. \label{p5}
\end{equation}

\section{Modular localization in complementary wedges}\label{sec:modular}

Let us consider a free massive scalar field in $1+1$ Minkowski spacetime and the complementary right and left wedges $W_R$ and $W_L$\footnote{The regions $W_R$ and $W_L$ are defined as:
\begin{equation} 
W_R =\{(t,x), \; x>|t| \} \;, \qquad W_L=\{(t,x), \; -x>|t| \}  \;. \label{wedges}
\end{equation}}. We denote the corresponding standard real one-particle subspaces by $\cM(W_R)$ and $\cM(W_L)=\cM'(W_R)$ \cite{Bisognano:1975ih,Brunetti:2002ygx,Guido:2008jk,Caribe:2026mod}., the prime standing for the symplectic complement. With the Bisognano-Wichmann normalization \cite{Bisognano:1975ih}, the one-particle modular operator and conjugation are
\begin{equation}
\delta = e^{-2\pi K} \;, \qquad j = \mathrm{CPT} \;, \qquad jKj = -K \;, \label{modBW}
\end{equation}
$K$ being the generator of the boosts, while the closed antilinear Tomita--Takesaki operators read
\begin{equation}
s = j \delta^{1/2}\;, \qquad s^\dagger = j \delta^{-1/2} \;, \label{tomita}
\end{equation}
their fixed-point spaces being $\cM(W_R)$ and $\cM(W_L)$, respectively \cite{Bisognano:1975ih,Brunetti:2002ygx,Guido:2008jk,Caribe:2026mod}.

\subsection{Spectral domains and strip analyticity}\label{domains}

Since the construction below relies on unbounded modular operators, it is convenient to make the domain conventions explicit. In rapidity space we choose
\begin{equation}
K = -i \partial_\theta \;, \qquad {\hat \psi}(\omega) = \frac{1}{\sqrt{2\pi}} \int_{{\mathbb R}} e^{-i \omega \theta} \psi(\theta) d\theta \;, \label{Kb}
\end{equation}
so that $K$ acts on ${\hat \psi}$ as the multiplication by $\omega$ and
\begin{equation}
{\widehat{\delta^a \psi}}(\omega) = e^{-2\pi a \omega} {\hat \psi}(\omega) \;, \qquad a \in {\mathbb R} \;. \nonumber
\end{equation}
Consequently
\begin{equation}
D(\delta^a) = \left\{ h \in L^2({\mathbb R}, d\omega) \;:\; \int_{{\mathbb R}} e^{-4\pi a \omega} |h(\omega)|^2 d\omega < \infty \right\} \;, \label{domd}
\end{equation}
and, for the Tomita--Takesaki operators,
\begin{eqnarray}
D(s) = D(\delta^{1/2}) & = & \left\{ h \;:\; \int_{{\mathbb R}} e^{-2\pi \omega} |h(\omega)|^2 d\omega < \infty \right\} \;, \nonumber \\
D(s^\dagger) = D(\delta^{-1/2}) & = & \left\{ h \;:\; \int_{{\mathbb R}} e^{2\pi \omega} |h(\omega)|^2 d\omega < \infty \right\} \;. \label{doms}
\end{eqnarray}
These are conditions of weighted square integrability. The weight $e^{2\pi \omega}$ appearing in a squared norm characterizes $D(s^\dagger)$, while multiplication by $e^{2\pi\omega}$ itself corresponds to $\delta^{-1}$.

The weighted $L^2$ conditions are the spectral form of strip analyticity. Formally,
\begin{equation}
\left( \delta^a \psi\right)(\theta) = \psi(\theta + 2\pi i a) \;, \label{imtr}
\end{equation}
More precisely, $D(\delta^{1/2})$ corresponds to the $L^2$ boundary values of functions analytic in the strip $0 < \Im z < \pi$, while $D(\delta)$ corresponds to a strip of width $2\pi$, the reflected strip describing $D(\delta^{-1/2})$. The opposite convention $K=+i\partial_\theta$ reverses the orientation of the strips.

Analyticity specifies the domain of the Tomita--Takesaki operator; localization is the additional fixed-point condition
\begin{equation}
{\cM(W_R)} = \left\{ \xi \in D(s) \;\; s \xi = \xi \right\} \;, \qquad {\cM(W_L)} = \left\{ \xi \in D(s^\dagger)\;:\; s^\dagger \xi = \xi \right\} \;. \label{fixp}
\end{equation}
The packet introduced below is a seed vector in $D(s) \cap D(s^\dagger)$. The combinations $(1+s)\Psi_d$ and $(1+s^\dagger)\Psi_d$, rather than $\Psi_d$ itself, are localized in the wedges.

\subsection{A normalizable finite-band packet}\label{packet}

Working in the spectral representation in which $K$ acts as the multiplication by $\omega$, we choose a normalized vector $\Psi_d$ with flat spectral density
\begin{equation}
h_d(\omega) = \frac{1}{\sqrt{2d}} \; \chi_{[\omega_0-d,\,\omega_0+d]}(\omega) \;, \qquad 0 < d < \omega_0 \;, \label{pack}
\end{equation}
its reflected vector $j\Psi_d$ having support in the corresponding negative band. Although $h_d$ has jump discontinuities, the compactness of its spectral support already implies $h_d \in D(\delta^a)$ for every $a \in {\mathbb R}$. Indeed, the inverse Fourier transform
\begin{equation}
\psi_d(z) = \frac{e^{i\omega_0 z}}{\sqrt{\pi d}} \; \frac{\sin(dz)}{z} = \frac{1}{\sqrt{2\pi}} \int d\omega \; e^{i \omega z} h_d(\omega) \;, \label{psid}
\end{equation}
whose singularity at $z=0$ is removable, is entire, and
\begin{equation}
||\psi_d(\,\cdot + iy\,)||_2^2 = \frac{1}{2d}\int_{\omega_0-d}^{\omega_0+d} e^{-2\omega y} d\omega = e^{-2\omega_0 y} \frac{\sinh(2dy)}{2dy} \;, \label{stripn}
\end{equation}
the last factor being understood as $1$ at $y=0$. Every finite strip therefore has finite $L^2$ norm. The endpoint jumps produce only a sinc-like decay along the real rapidity axis; strip analyticity depends on exponential spectral moments, not on pointwise continuity in $\omega$.

Let us introduce
\begin{equation}
\lambda^2 = e^{-2\pi \omega_0} \;, \qquad S_d = \frac{\sinh(2\pi d)}{2\pi d} \;, \qquad \mu_+ = \lambda^2 S_d \;, \qquad \mu_- = \lambda^{-2} S_d \;, \label{lamS}
\end{equation}
so that $\mu_+ < 1 < \mu_-$, the whole band lying in $(0,\infty)$. We then define
\begin{eqnarray}
f & = & (1-\mu_+)^{-1/2} (1+s)\Psi_d \;, \qquad \;\;\; f' = (1-\mu_+)^{-1/2}(1+s)\, i \Psi_d \;, \nonumber \\
g & = & (\mu_- -1)^{-1/2} (1+s^\dagger)\Psi_d \;, \qquad g' = -(\mu_- -1)^{-1/2}(1+s^\dagger)\, i\Psi_d \;, \label{fgpair}
\end{eqnarray}
so that $(f,f') \in \cM(W_R)$ and $(g,g') \in \cM(W_L)$. A direct computation, given in Appendix~\ref{appGram}, gives the inner products
\begin{eqnarray}
||f||^2 = ||f'||^2 & = & \nu_A = \frac{1+\mu_+}{1-\mu_+} \;, \qquad ||g||^2 = ||g'||^2 = \nu_B = \frac{1+\mu_-}{\mu_- -1} \;, \nonumber \\
\langle f|f'\rangle & = & i \;, \qquad \langle g|g'\rangle = i \;, \nonumber \\
\langle f|g\rangle & = & \kappa \;, \qquad \langle f'|g'\rangle = -\kappa \;, \nonumber \\
\langle f|g'\rangle & = & \langle f'|g\rangle = 0 \;, \label{gram}
\end{eqnarray}
with
\begin{equation}
\kappa = \frac{2}{\sqrt{(1-\mu_+)(\mu_- -1)}} \;. \label{kappa}
\end{equation}
The symplectic normalization within each wedge,
$\Delta_{PJ}(f,f')=\Delta_{PJ}(g,g')=2$, is exact at nonvanishing bandwidth. All cross products are real, as required by the space-like separation of the wedges, and $\langle f|g\rangle=\kappa$ is their Hadamard correlation.

The flat density \eqref{pack} was chosen because its modular moments are elementary. It may be replaced by a real normalized amplitude $h \in C_0^\infty((\omega_0-d,\omega_0+d))$, with the substitutions
\begin{equation}
\mu_+(h) = \int_{{\mathbb R}} e^{-2\pi \omega} |h(\omega)|^2 d\omega \;, \qquad \mu_-(h) = \int_{{\mathbb R}} e^{2\pi\omega} |h(\omega)|^2 d\omega \;. \label{smooth}
\end{equation}
Smooth amplitudes supported in the same positive band approximate $h_d$ in $L^2$, and their weighted moments, inner products and Bell-CHSH correlators converge to those of the flat packet. Every strict violation therefore persists under a sufficiently accurate spectral smoothing. 
\subsection{The narrow-band limit}\label{narrow}

When $d \to 0$ at fixed $\omega_0$, one has $\mu_+ \to \lambda^2$ and $\mu_- \to \lambda^{-2}$, so that the inner products \eqref{gram} reduce to
\begin{equation}
\nu_A , \nu_B \;\longrightarrow\; \nu_\lambda = \frac{1+\lambda^2}{1-\lambda^2} \;, \qquad \kappa \;\longrightarrow\; \kappa_\lambda = \frac{2\lambda}{1-\lambda^2} \;, \label{ideal}
\end{equation}
which are the familiar two-mode modular expressions used in the Summers-Werner construction \cite{Summers:1987squ} and in its rapidity-space realization \cite{Caribe:2026mod}. At nonvanishing $d$, the vector $\Psi_d$ is normalizable and eqs.~\eqref{gram} are exact; all numerical values below are evaluated at finite bandwidth.

Writing
\begin{equation}
\lambda = \tanh r \;, \qquad r > 0 \;, \label{lamr}
\end{equation}
expressions \eqref{ideal} become
\begin{equation}
\nu_\lambda = \frac{1+\tanh^2 r}{1-\tanh^2 r} = \cosh(2r) \;, \qquad \kappa_\lambda = \frac{2\tanh r}{1-\tanh^2 r} = \sinh(2r) \;, \label{tmsv}
\end{equation}
which are the covariances of a two-mode squeezed vacuum of squeezing parameter $r$. Thus the finite-dimensional quasifree covariance converges to the two-mode squeezed covariance, with the edge $\lambda \to 1$ corresponding to infinite squeezing. This supplies a useful check of the formulas without changing the fact that all values reported below use normalizable finite-band packets.

\section{Violation of the Bell-CHSH inequality and the approach to Tsirelson's bound}\label{sec:construction}

\subsection{The four settings}\label{settings}

We are now ready to assemble the Bell-CHSH operator. Let $p$ be any polynomial of the form \eqref{pdef} with $||p||_\infty \le 1$, and let $(f,f',g,g')$ be the modular vectors \eqref{fgpair}. Alice's operators are taken to be the two axis observables in the right wedge,
\begin{equation}
A(f) = S_p(f) \;, \qquad A(f') = S_p(f') \;, \label{Aset}
\end{equation}
while Bob's operators are the two combinations \eqref{BBp} of the axis observables in the left wedge,
\begin{equation}
B(g) = \frac{S_p(g)+S_p(g')}{\sqrt{2}} \;, \qquad B(g') = \frac{S_p(g)-S_p(g')}{\sqrt{2}} \;. \label{Bset}
\end{equation}
Alice's and Bob's operators commute, the two wedges being space-like separated. Within each wedge, eqs.\eqref{gram} give $\Delta_{PJ}(f,f')=\Delta_{PJ}(g,g')=2$, so that
\begin{equation}
\{ S_p(f), S_p(f')\} = 0 \;, \qquad \{ S_p(g), S_p(g')\}=0 \;. \label{news}
\end{equation}
and therefore $||A(f)||,||A(f')||,||B(g)||,||B(g')|| \le 1$. Notice also that the bounds $\|A(f)\|,\|A(f')\|\leq 1$ follow directly
from $\|p\|_\infty\leq 1$ and the spectral theorem, independently of
the anticommutation relation \eqref{news}; the latter is needed for the
corresponding normalization of Bob's combinations $B(g)$ and $B(g')$.The four settings \eqref{Aset},\eqref{Bset} are finite Weyl polynomials localized in complementary wedges. Substituting them into \eqref{CHSHop}, the four cross terms cancel pairwise and one is left with
\begin{equation}
{\cC} = \sqrt{2}\left[ S_p(f) S_p(g) + S_p(f')S_p(g') \right] \;. \label{Bred}
\end{equation}
Equation \eqref{Bred} displays a useful structural simplification of the
Bell-CHSH operator. Owing to the particular choice of Bob's settings
in Eq.\eqref{Bset}, the four terms of the original CHSH combination reorganize
themselves into only two diagonal products, while the mixed terms cancel
exactly. Thus the Bell problem is reduced to the evaluation of the two
correlators associated with the pairs $(f,g)$ and $(f',g')$. Moreover,
the modular construction makes these two contributions equivalent in the
vacuum, so that the expectation value of the full CHSH operator is
ultimately determined by a single finite-polynomial correlator, as shown
below.

\subsection{The vacuum correlator}\label{correl}

Expanding each cosine into two Weyl operators according to \eqref{SWeyl} and using \eqref{vacW} together with the inner products \eqref{gram}, the vacuum correlator is obtained in closed form as a finite double sum,
\begin{eqnarray}
E_p(\lambda,d) & \equiv & \langle 0|\; S_p(f) S_p(g)\;|0\rangle \nonumber \\
& = & \frac{1}{2} \sum_{m,n \in M} c_m c_n \left[ \exp\left\{ -\frac{\alpha^2}{2}\left( m^2 \nu_A + n^2 \nu_B + 2mn\kappa\right) \right\} \right. \nonumber \\
& & \hspace{25mm} \left. + \exp\left\{ -\frac{\alpha^2}{2}\left( m^2\nu_A + n^2\nu_B - 2mn\kappa \right) \right\} \right] \;. \label{Efin}
\end{eqnarray}
The expression is invariant under $\kappa \to -\kappa$, so that $\langle 0|S_p(f')S_p(g')|0\rangle = E_p(\lambda,d)$ as well, and, from \eqref{Bred},
\begin{equation}
\langle 0|\; {\cC}\;|0\rangle = 2\sqrt{2} \; E_p(\lambda,d) \;. \label{CHSHfin}
\end{equation}
Equation \eqref{CHSHfin} is the main tool of the present investigation. It is exact, it involves only finitely many terms, and every quantity entering it, $(\nu_A,\nu_B,\kappa)$, is given in closed form by eqs.\eqref{lamS},\eqref{gram},\eqref{kappa} in terms of the two packet parameters $(\lambda,d)$.

\subsection{An economical violation}\label{econ}

Let us first look at the three-harmonic polynomial \eqref{psix}. As shown in Sect.~\ref{approach}, the limiting value of \eqref{CHSHfin} for a fixed polynomial is $\sqrt{2}\sum_m c_m^2$, which here gives
\begin{equation}
\sqrt{2} \sum_{m \in \{1,3,5\}} c_m^2 = \frac{77\sqrt{2}}{50} = 2.1778888861 \;. \label{sixlim}
\end{equation}
Already at the exact normalizable finite-band point
\begin{equation}
\lambda = 0.99 \;, \qquad d = 0.001 \;, \qquad \omega_0 = 0.003199121262 > d \;, \nonumber
\end{equation}
the direct substitution in \eqref{Efin},\eqref{CHSHfin} gives
\begin{equation}
\langle 0|\;{\cC}\;|0\rangle = 2.1488540073 \;, \label{sixval}
\end{equation}
which is a violation obtained with neither a zero-width packet nor an improper modular eigenvector. Each of Alice's operators contains six Weyl operators and, after expanding \eqref{Bset}, each of Bob's operators contains twelve. For the normalizable point $(\lambda,d)=(0.999,10^{-4})$ the same polynomial gives $2.1749324922$.

\subsection{The approach to Tsirelson's bound}\label{approach}

For fixed $r \in (0,1)$, let $\mathfrak{F}_r$ denote the family of Bell-CHSH operators obtained from the normalized Fejér polynomials $p_L$ of eq.\eqref{pL} and from the compact bands $d = r\omega_0$, with $L < \infty$ and $\omega_0 > 0$. We want to show that
\begin{equation}
\sup_{{\cC} \in \mathfrak{F}_r} \langle 0|\;{\cC}\;|0\rangle = 2\sqrt{2} \;, \label{sup}
\end{equation}
no finite member of the family attaining the supremum, so that for every $\varepsilon>0$ there is a finite quadruple in $\mathfrak{F}_r$ with
\begin{equation}
2\sqrt{2} - \varepsilon < \langle 0|\;{\cC}\;|0\rangle < 2\sqrt{2} \;. \label{eps}
\end{equation}
Let us set $\lambda^2 = e^{-2\pi\omega_0}$, $d = r\omega_0$, and let $\omega_0 \downarrow 0$. Every packet of the sequence is normalizable and has strictly positive bandwidth. From eqs.\eqref{gram},\eqref{kappa},
\begin{equation}
\nu_A , \nu_B , \kappa = \frac{1}{\pi \omega_0} + O(1) \;, \qquad \nu_A + \nu_B - 2\kappa \;\longrightarrow\; 0 \;, \nonumber
\end{equation}
the exact positive-square identity and the required off-diagonal asymptotics being derived in Appendix~\ref{appZero}. In the double sum \eqref{Efin}, the exponential carrying the positive cross term therefore vanishes for every pair $(m,n)$, while the one carrying the negative cross term vanishes for $m \neq n$ and tends to $1$ for $m=n$. Hence, for every fixed finite polynomial,
\begin{equation}
\lim_{\substack{\omega_0 \downarrow 0 \\ d = r\omega_0}} \langle 0|\;{\cC}\;|0\rangle = \sqrt{2} \sum_{m \in M} c_m^2 \;, \label{fixplim}
\end{equation}
which is the value quoted in \eqref{sixlim}. In particular, a fixed bounded odd-harmonic polynomial produces a strict asymptotic violation whenever $\sum_m c_m^2 > \sqrt{2}$. Taking now $p=p_L$ and using the third of eqs.\eqref{pnorm},
\begin{equation}
\lim_{L\to\infty} \; \lim_{\substack{\omega_0 \downarrow 0 \\ d=r\omega_0}} \langle 0|\;{\cC}\;|0\rangle = 2\sqrt{2} \;, \nonumber
\end{equation}
which establishes the lower bound in \eqref{sup}, the upper bound being provided by \eqref{contr}.

It remains to check that the supremum is never attained by a finite member. Introducing the vacuum mean squares
\begin{equation}
s_A = \langle 0|\;S_p(f)^2\;|0\rangle \;, \qquad s_B = \langle 0|\;S_p(g)^2\;|0\rangle \;, \label{sAsB}
\end{equation}
and using the anticommutation $\{S_p(g),S_p(g')\}=0$, which gives $\langle 0|B(g)^2|0\rangle = \langle 0|B(g')^2|0\rangle = s_B$, the contraction bound \eqref{contr} reads
\begin{equation}
\left| \langle 0|\;{\cC}\;|0\rangle \right| \le 2 \sqrt{2\, s_A s_B} \;. \label{contr2}
\end{equation}
For every nonvanishing $h$, the vacuum spectral distribution of $\varphi(h)$ is a nondegenerate Gaussian with full support, while $p$ is continuous, nonconstant and bounded by one in modulus, so that
\begin{equation}
s_A = \int_{{\mathbb R}} p(\alpha x)^2 \, d\gamma_{\nu_A}(x) < 1 \;, \nonumber
\end{equation}
and likewise for $s_B$. Hence $|\langle 0|{\cC}|0\rangle| < 2\sqrt{2}$ at every finite bandwidth, which is the strict upper inequality in \eqref{eps}.

A direct Weyl expansion also gives
\begin{equation}
s_A = \frac{1}{2}\sum_{m,n \in M} c_m c_n \left[ e^{-\frac{\alpha^2}{2}(m-n)^2\nu_A} + e^{-\frac{\alpha^2}{2}(m+n)^2 \nu_A}\right] \;\xrightarrow[\;\nu_A \to \infty\;]{}\; \frac{1}{2}\sum_{m \in M}c_m^2 \;, \label{sAlimit}
\end{equation}
and the same for $s_B$. In the modular concentration limit $\omega_0 \downarrow 0$ one therefore has
\begin{equation}
2\sqrt{2\,s_A s_B} \;\longrightarrow\; \sqrt{2} \sum_{m \in M} c_m^2 \;, \nonumber
\end{equation}
which coincides with the limiting value \eqref{fixplim} of the correlator. Thus, along the modular-concentration path, the construction asymptotically saturates the corresponding contraction bound. The remaining deficit in this ordered limit is the mean-square deficit $1-\frac{1}{2}\sum_m c_m^2$ of the polynomial. The Parseval bound \eqref{parseval} shows that this deficit is positive at every finite degree and vanishes for the Fejér sequence.

The convergence is in the vacuum mean square, not in the operator norm: the right hand side of \eqref{sAlimit} tends to $1$ as $L \to \infty$, while $p_L(\pi/2)=0$ for every $L$.

As a representative near-bound finite point, let us take
\begin{equation}
L = 256 \;, \qquad (\lambda,d) = (0.99999,10^{-6})\;, \qquad \omega_0 = 3.1831147774 \times 10^{-6} > d \;. \nonumber
\end{equation}
The degree-$511$ polynomial contains $512$ Weyl operators per axis observable and the exact finite sum \eqref{Efin} gives
\begin{equation}
\langle 0|\;{\cC}\;|0\rangle = 2.8002741385 \;, \label{near}
\end{equation}
to be compared with the spectral limit \eqref{fixplim}, which equals $2.8017677455$. At the above parameters one has
\begin{equation}
\nu_A = 99999.53 \;, \qquad \nu_B = 99999.47\;, \qquad \kappa = 99999.50 \;, \nonumber
\end{equation}
so that the packet is strongly concentrated near the modular fixed point $\omega=0$. Approaching $2\sqrt{2}$ requires both a high-degree polynomial and large one-particle norms. The wedge localization remains exact at every stage. Further finite-band values are collected in Table~\ref{tab:Fejer}.
\begin{table}[h!]
\centering
\begin{tabular}{ccccc}
\toprule
$L$ & degree & Weyl operators per axis & spectral limit & finite-band value \\
\midrule
5  & 9  & 10 & 2.2947679777 & 2.2911615070 \\
10 & 19 & 20 & 2.4953473110 & 2.4886420495 \\
16 & 31 & 32 & 2.5910819755 & 2.5808459643 \\
32 & 63 & 64 & 2.6871179712 & 2.6689354013 \\
\bottomrule
\end{tabular}
\caption{Exact finite-band Bell-CHSH values for the normalized Fejér polynomials, at $\lambda=0.999$, $d=10^{-4}$. Each of Bob's operators contains twice the displayed number of Weyl operators. No entry makes use of an improper or zero-width packet.}
\label{tab:Fejer}
\end{table}

It is worth emphasizing the modular-theoretic meaning of the limit
displayed in Table~I. Since
\begin{equation}
 \lambda^2=e^{-2\pi\omega_0},
\end{equation}
the limit $\lambda\to1$ is precisely the limit $\omega_0\to0$, namely
the concentration of the spectral packet near the zero of the modular
(boost) generator. It is in this regime that the Bell-CHSH
correlator approaches its maximal value. This observation is closely
related to the modular picture underlying the Summers-Werner results
on maximal Bell correlations in relativistic quantum field theory
\cite{Summers:1985wq,Summers:1987fn,Summers:1987squ,Summers:1987ze}. In particular, under the
standard assumptions considered there, the von Neumann algebras
associated with wedge regions are factors of type~$\mathrm{III}_1$
and exhibit maximal Bell correlations. From this perspective, the
limit $\lambda\to1$ appearing in the present explicit construction is
not merely a convenient numerical regime: it singles out the modular
spectral region near eigenvalue one, or equivalently near zero modular
frequency, which is naturally associated with the type~$\mathrm{III}_1$
structure of local quantum field theory.



\section{Conclusion}\label{concl}

We have constructed four bounded Hermitian operators, each one a finite polynomial in the unitary Weyl operators, which violate the Bell-CHSH inequality in the vacuum of a massive scalar field. The construction uses odd Weyl harmonics and symplectically normalized modular vectors. The two axis observables in each wedge anticommute exactly, while Bob's final settings are their normalized sum and difference.

The finite-band packet gives all one-particle inner products in closed form. The Bell-CHSH correlator is consequently reduced to the finite sum \eqref{Efin}. Six Weyl terms per axis already give $2.14885$. For the normalized Fejér family, the ordered modular-concentration and large-degree limits give
\begin{equation}
\sup_{{\cC}\in\mathfrak F_r}\langle0|{\cC}|0\rangle=2\sqrt{2}\;,
\end{equation}
although every finite member stays strictly below the supremum. The degree-$511$ example gives $2.80027$.

The Gaussian observation of Sect.~\ref{gauss} explains why the final settings matter. Four final settings that are bounded functions of individual quadratures admit one common positive Gaussian representation in a centered quasifree state. Our construction leaves this class because each Bob setting mixes two noncommuting axes. The Hadamard correlations remain present throughout and determine the cross-wedge quantity $\kappa$.

The construction is exact in the wedge algebras and in the boost-spectral representation. By the standardness and density properties of the one-particle
localization subspaces, the modular vectors used here can be
approximated by vectors generated by compactly supported test
functions in the corresponding wedges; see Ref.\cite{Guido:2008jk}. Since the
Weyl representation is strongly continuous, every strict
Bell-CHSH violation persists under a sufficiently accurate
such approximation.

Both the polynomial degree and the one-particle norms grow as the
correlator approaches $2\sqrt{2}$. Understanding whether the same
asymptotic behavior can be achieved with more economical
finite-polynomial families provides a natural direction for further
investigation. As such, it would be worth exploring alternative harmonic
regularizations of the square-wave observable
$q(x)=\operatorname{sgn}(\cos x)$. As briefly discussed in
Appendix \eqref{app:abel-poisson} Abel-Poisson summation provides a particularly simple
example: it preserves the odd-harmonic structure required for exact
anticommutation, while damping the Fourier modes geometrically through
factors $r^m$. Finite truncations therefore lead to another family of
finite Weyl-polynomial observables, after the appropriate norm
normalization. It would be interesting to determine whether optimized
Abel-Poisson polynomials, or related summability schemes, can provide
a more efficient approach to maximal Bell-CHSH violation at fixed
polynomial degree.

\section*{Acknowledgments}
The authors would like to thank the Brazilian agencies Conselho Nacional de Desenvolvimento Científico e Tecnológico (CNPq), Coordenação de Aperfeiçoamento de Pessoal de Nível Superior $-$ Brasil (CAPES) and Fundação Carlos Chagas Filho de Amparo à Pesquisa do Estado do Rio de Janeiro (FAPERJ) for financial support. In particular, S.~P.~Sorella, I.~Roditi, and M.~S.~Guimaraes are CNPq researchers under contracts 301030/2019-7, 319060/2025-0,  and 309793/2023-8, respectively.

\section*{Declaration on the use of generative AI}
During the preparation of this manuscript, the authors used OpenAI GPT-5.6 Sol and Anthropic Claude Opus 5 as assistive tools for drafting and revising the text, reorganizing the presentation, checking algebraic and numerical calculations, and supporting the verification of references. All arguments, calculations, citations, and conclusions were subsequently reviewed by the authors. The authors take full responsibility for the scientific content and final form of the manuscript.

\section*{Data availability}
The data generated in this article can be provided upon reasonable request.

\appendix

\section{The finite-band inner products}\label{appGram}

Let us derive eqs.\eqref{gram},\eqref{kappa} in the boost-spectral representation. The one-particle Hilbert space is identified with
\begin{equation}
{\cal H}_1 \simeq L^2({\mathbb R},d\omega) \;, \qquad
\langle \psi_1|\psi_2\rangle
=
\int_{\mathbb R}\overline{\psi_1(\omega)}\,\psi_2(\omega)\,d\omega \;, \label{appinner}
\end{equation}
where the inner product is linear in its second entry. The variable $\omega$ is the spectral variable of the boost generator $K$, or equivalently the Fourier variable conjugate to the rapidity $\theta$ in eq.\eqref{Kb}; it is not the Minkowski energy. In this representation we identify the vector $\Psi_d$ with its spectral wave function,
\begin{equation}
\Psi_d(\omega)=h_d(\omega)
=
\frac{1}{\sqrt{2d}}\,\chi_{I_+}(\omega) \;, \qquad
I_+=[\omega_0-d,\omega_0+d] \;. \label{apppacket}
\end{equation}
The function $\psi_d(z)$ in eq.\eqref{psid} is the inverse Fourier transform of this spectral wave function. Since $0<d<\omega_0$, one has
\begin{equation}
I_+\subset(0,\infty) \;, \qquad
I_-=-I_+=[-\omega_0-d,-\omega_0+d]\subset(-\infty,0) \;, \qquad
I_+\cap I_-=\varnothing \;. \label{bands}
\end{equation}
Equation \eqref{apppacket} also gives $||\Psi_d||^2=1$.

Let us now state explicitly the modular action used below. With the phase convention adopted for the scalar one-particle representation,
\begin{equation}
(K\psi)(\omega)=\omega\psi(\omega) \;, \qquad
(\delta^a\psi)(\omega)=e^{-2\pi a\omega}\psi(\omega) \;, \qquad
(j\psi)(\omega)=\overline{\psi(-\omega)} \;, \label{appactions}
\end{equation}
where $a\in{\mathbb R}$. Thus $j$ is antiunitary, $j^2=1$, and $jKj=-K$. The complex conjugation in \eqref{appactions} is required by antiunitarity. Since $h_d$ is real, it does not appear explicitly when $j$ acts on the packet itself:
\begin{eqnarray}
(j\Psi_d)(\omega) & = & h_d(-\omega) \;, \nonumber\\
(j\delta^a\Psi_d)(\omega) & = & e^{2\pi a\omega}h_d(-\omega) \;. \label{appjd}
\end{eqnarray}
The first function is supported in $I_-$, and multiplication by the exponential in the second line does not change that support. In particular,
\begin{equation}
\operatorname{supp}(\delta^a\Psi_d)=I_+ \;, \qquad
\operatorname{supp}(j\delta^a\Psi_d)=I_- \;. \nonumber
\end{equation}

The orthogonality used in the calculation can now be seen directly. For every real $a$,
\begin{eqnarray}
\langle\Psi_d|j\delta^a\Psi_d\rangle
&=&
\int_{\mathbb R}h_d(\omega)e^{2\pi a\omega}h_d(-\omega)\,d\omega \nonumber\\
&=&
\frac{1}{2d}\int_{\mathbb R}e^{2\pi a\omega}
\chi_{I_+}(\omega)\chi_{I_-}(\omega)\,d\omega
=0 \;, \label{apporth}
\end{eqnarray}
because the two bands in \eqref{bands} are disjoint. Taking $a=1$ gives explicitly $\langle\Psi_d|j\delta\Psi_d\rangle=0$, while $a=\pm1/2$ gives the mixed terms involving $s=j\delta^{1/2}$ and $s^\dagger=j\delta^{-1/2}$. Their conjugates vanish as well.

From eqs.\eqref{tomita},\eqref{appactions}, the two Tomita--Takesaki operators act on the packet as
\begin{equation}
(s\Psi_d)(\omega)=e^{\pi\omega}h_d(-\omega) \;, \qquad
(s^\dagger\Psi_d)(\omega)=e^{-\pi\omega}h_d(-\omega) \;. \label{appst}
\end{equation}
Both vectors are supported in $I_-$. Their norms are
\begin{eqnarray}
||s\Psi_d||^2
&=&
\frac{1}{2d}\int_{I_-}e^{2\pi\omega}\,d\omega
=
\frac{1}{2d}\int_{\omega_0-d}^{\omega_0+d}e^{-2\pi u}\,du
=
\lambda^2\frac{\sinh(2\pi d)}{2\pi d}
=\mu_+ \;, \nonumber\\
||s^\dagger\Psi_d||^2
&=&
\frac{1}{2d}\int_{I_-}e^{-2\pi\omega}\,d\omega
=
\frac{1}{2d}\int_{\omega_0-d}^{\omega_0+d}e^{2\pi u}\,du
=
\lambda^{-2}\frac{\sinh(2\pi d)}{2\pi d}
=\mu_- \;, \label{appnorms}
\end{eqnarray}
where $u=-\omega$ was used in the second form of each line. A further inner product needed below is
\begin{eqnarray}
\langle s\Psi_d|s^\dagger\Psi_d\rangle
&=&
\int_{\mathbb R}
\overline{e^{\pi\omega}h_d(-\omega)}\,
e^{-\pi\omega}h_d(-\omega)\,d\omega \nonumber\\
&=&
\int_{\mathbb R}|h_d(-\omega)|^2d\omega
=1 \;. \label{appcrossseed}
\end{eqnarray}
This is also an immediate consequence of the antiunitarity of $j$:
$\langle j\delta^{1/2}\Psi_d|j\delta^{-1/2}\Psi_d\rangle
=\langle\delta^{-1/2}\Psi_d|\delta^{1/2}\Psi_d\rangle=1$.

For clarity, let us abbreviate
\begin{equation}
u=\Psi_d \;, \qquad v=s\Psi_d \;, \qquad w=s^\dagger\Psi_d \;, \qquad
N_A=(1-\mu_+)^{-1/2} \;, \qquad N_B=(\mu_- - 1)^{-1/2} \;. \nonumber
\end{equation}
Equations \eqref{apporth}-\eqref{appcrossseed} give
\begin{equation}
\langle u|u\rangle=1 \;, \quad
\langle v|v\rangle=\mu_+ \;, \quad
\langle w|w\rangle=\mu_- \;, \quad
\langle u|v\rangle=\langle u|w\rangle=0 \;, \quad
\langle v|w\rangle=1 \;, \label{appbasic}
\end{equation}
together with the conjugate relations. Since $s$ and $s^\dagger$ are antilinear,
\begin{equation}
(1+s)i\Psi_d=i(u-v) \;, \qquad
(1+s^\dagger)i\Psi_d=i(u-w) \;. \nonumber
\end{equation}
The four vectors in eq.\eqref{fgpair} can therefore be written as
\begin{equation}
f=N_A(u+v) \;, \qquad
f'=iN_A(u-v) \;, \qquad
g=N_B(u+w) \;, \qquad
g'=-iN_B(u-w) \;. \label{appfg}
\end{equation}

We first compute the inner products within the right wedge. Using \eqref{appbasic},
\begin{eqnarray}
\langle u+v|u+v\rangle &=& 1+\mu_+ \;, \nonumber\\
\langle i(u-v)|i(u-v)\rangle &=& 1+\mu_+ \;, \nonumber\\
\langle u+v|i(u-v)\rangle &=& i(1-\mu_+) \;. \nonumber
\end{eqnarray}
Multiplication by $N_A^2$ gives
\begin{equation}
||f||^2=||f'||^2=\frac{1+\mu_+}{1-\mu_+} \;, \qquad
\langle f|f'\rangle=i \;. \label{appAlice}
\end{equation}
Similarly,
\begin{eqnarray}
\langle u+w|u+w\rangle &=& 1+\mu_- \;, \nonumber\\
\langle -i(u-w)|-i(u-w)\rangle &=& 1+\mu_- \;, \nonumber\\
\langle u+w|-i(u-w)\rangle &=& i(\mu_- - 1) \;, \nonumber
\end{eqnarray}
and multiplication by $N_B^2$ yields
\begin{equation}
||g||^2=||g'||^2=\frac{1+\mu_-}{\mu_- - 1} \;, \qquad
\langle g|g'\rangle=i \;. \label{appBob}
\end{equation}

It remains to evaluate the cross-wedge products. The terms containing one of $v,w$ and one $u$ vanish by \eqref{apporth}, whereas $\langle u|u\rangle=\langle v|w\rangle=1$. Consequently,
\begin{eqnarray}
\langle f|g\rangle
&=&N_AN_B\left(\langle u|u\rangle+\langle v|w\rangle\right)
=2N_AN_B
=\frac{2}{\sqrt{(1-\mu_+)(\mu_- - 1)}}
=\kappa \;, \nonumber\\
\langle f'|g'\rangle
&=&-N_AN_B\left(\langle u|u\rangle+\langle v|w\rangle\right)
=-\kappa \;, \nonumber\\
\langle f|g'\rangle
&=&-iN_AN_B\left(\langle u|u\rangle-\langle v|w\rangle\right)
=0 \;, \nonumber\\
\langle f'|g\rangle
&=&-iN_AN_B\left(\langle u|u\rangle-\langle v|w\rangle\right)
=0 \;. \label{appcrossfinal}
\end{eqnarray}
Equations \eqref{appAlice},\eqref{appBob},\eqref{appcrossfinal} reproduce eqs.\eqref{gram},\eqref{kappa} and complete the finite-band calculation.

\section{The modular zero-spectrum limit}\label{appZero}

Setting $x=2\pi\omega_0$ and $d=r\omega_0$ with $0<r<1$, one has
\begin{equation}
S_d = \frac{\sinh(rx)}{rx} = 1 + \frac{r^2x^2}{6} + O(x^4) \;, \nonumber
\end{equation}
and therefore
\begin{eqnarray}
\mu_+ & = & 1 - x + \left( \frac{1}{2}+\frac{r^2}{6}\right)x^2 + O(x^3) \;, \nonumber \\
\mu_- & = & 1 + x + \left( \frac{1}{2}+\frac{r^2}{6}\right)x^2 + O(x^3) \;. \nonumber
\end{eqnarray}
Writing $u=1-\mu_+$ and $v=\mu_- -1$, the finite-band inner products become
\begin{equation}
\nu_A = \frac{2}{u}-1 \;, \qquad \nu_B = \frac{2}{v}+1 \;, \qquad \kappa = \frac{2}{\sqrt{uv}} \;, \nonumber
\end{equation}
so that the diagonal combination is the exact positive square
\begin{equation}
\nu_A + \nu_B - 2\kappa = 2\left( \frac{1}{\sqrt u} - \frac{1}{\sqrt v}\right)^2 = 2 \left( \frac{1}{2}+\frac{r^2}{6}\right)^2 x + O(x^2) \;, \label{diag}
\end{equation}
which vanishes at every fixed $r \in (0,1)$. For arbitrary positive integers $m,n$,
\begin{eqnarray}
m^2 \nu_A + n^2 \nu_B - 2mn\kappa & = & 2\left( \frac{m}{\sqrt u} - \frac{n}{\sqrt v}\right)^2 + n^2 - m^2 \nonumber \\
& = & \frac{2(m-n)^2}{x} + O(1) \qquad (m \neq n) \;, \label{offdiag}
\end{eqnarray}
while the combination with $+2mn\kappa$ diverges as $2(m+n)^2/x + O(1)$. Equations \eqref{diag},\eqref{offdiag} establish all the exponential limits used in \eqref{fixplim}. In particular, no additional assumption $d/\omega_0 \to 0$ is needed.

\section{Bounds on the trigonometric polynomials}\label{appPoly}

`

Before proving the specific bounds used in the main text, we briefly
recall the definition and the basic properties of Fej\'er approximants
that are relevant for our construction.

Let $f$ be a $2\pi$-periodic integrable function with Fourier series
\begin{equation}
 f(x)\sim \sum_{n\in\mathbb Z}\widehat f_n e^{inx},
 \qquad
 \widehat f_n=
 \frac{1}{2\pi}\int_{-\pi}^{\pi}f(t)e^{-int}\,dt .
 \label{eq:appC-fourier}
\end{equation}
The $N$-th partial Fourier sum is
\begin{equation}
 S_N f(x)=\sum_{|n|\leq N}\widehat f_n e^{inx}.
 \label{eq:appC-partial}
\end{equation}
The Fej\'er mean of order $N$ is defined as the arithmetic mean of the
first $N+1$ Fourier partial sums,
\begin{equation}
 \sigma_N f(x)
 =
 \frac{1}{N+1}\sum_{j=0}^{N}S_j f(x).
 \label{eq:appC-fejer-def}
\end{equation}
Equivalently,
\begin{equation}
 \sigma_N f(x)
 =
 \sum_{|n|\leq N}
 \left(1-\frac{|n|}{N+1}\right)
 \widehat f_n e^{inx}.
 \label{eq:appC-fejer-fourier}
\end{equation}
Thus, relative to an ordinary Fourier truncation, the Fourier
coefficients acquire the triangular Fej\'er weights
\begin{equation}
 1-\frac{|n|}{N+1}.
\end{equation}

The Fej\'er mean can also be written as a convolution,
\begin{equation}
 \sigma_N f(x)
 =
 \frac{1}{2\pi}
 \int_{-\pi}^{\pi}
 F_N(x-t)f(t)\,dt ,
 \label{eq:appC-fejer-convolution}
\end{equation}
where
\begin{equation}
 F_N(t)
 =
 \frac{1}{N+1}
 \left[
 \frac{\sin\!\left((N+1)t/2\right)}
      {\sin(t/2)}
 \right]^2
 \label{eq:appC-fejer-kernel}
\end{equation}
is the Fej\'er kernel. Its two basic properties are
\begin{equation}
 F_N(t)\geq 0,
 \qquad
 \frac{1}{2\pi}\int_{-\pi}^{\pi}F_N(t)\,dt=1.
 \label{eq:appC-kernel-properties}
\end{equation}
Consequently, if $|f(x)|\leq 1$ almost everywhere, then
\begin{equation}
 |\sigma_N f(x)|\leq 1.
 \label{eq:appC-contraction}
\end{equation}

Fej\'er's theorem states that, at every point $x$ at which the
one-sided limits exist,
\begin{equation}
 \sigma_N f(x)
 \longrightarrow
 \frac{f(x+0)+f(x-0)}{2}.
 \label{eq:appC-fejer-theorem}
\end{equation}
In particular, at every continuity point of $f$,
\begin{equation}
 \sigma_N f(x)\longrightarrow f(x).
 \label{eq:appC-continuity}
\end{equation}

For the application considered in this work we take
\begin{equation}
 q(x)=\operatorname{sgn}(\cos x),
 \label{eq:appC-q}
\end{equation}
whose Fourier series contains only odd cosine harmonics,
\begin{equation}
 q(x)
 =
 \frac{4}{\pi}
 \sum_{k=0}^{\infty}
 \frac{(-1)^k}{2k+1}
 \cos\bigl((2k+1)x\bigr).
 \label{eq:appC-q-fourier}
\end{equation}
Choosing $N=2L-1$, its Fej\'er mean is therefore
\begin{equation}
 q_L(x)
 \equiv
 \sigma_{2L-1}q(x)
 =
 \frac{4}{\pi}
 \sum_{k=0}^{L-1}
 \frac{(-1)^k}{2k+1}
 \left(
 1-\frac{2k+1}{2L}
 \right)
 \cos\bigl((2k+1)x\bigr).
 \label{eq:appC-qL}
\end{equation}
Thus the Fej\'er procedure preserves the odd-harmonic structure which
is essential for the exact anticommutation argument of Sect.~III.

Since $|q(x)|=1$ almost everywhere and the Fej\'er kernel is positive
and normalized, one immediately obtains
\begin{equation}
 |q_L(x)|\leq 1.
 \label{eq:appC-qL-bound}
\end{equation}
For the particular function $q(x)=\operatorname{sgn}(\cos x)$, a
stronger statement holds: the maximum modulus of $q_L$ occurs at
$x=0$. Defining
\begin{equation}
 M_L\equiv q_L(0),
 \label{eq:appC-ML}
\end{equation}
one has
\begin{equation}
 \|q_L\|_\infty=M_L.
 \label{eq:appC-qL-norm}
\end{equation}
This property will be established explicitly below.

Since $q$ is continuous at $x=0$ and $q(0)=1$, Fej\'er's theorem gives
\begin{equation}
 M_L\longrightarrow 1,
 \qquad L\longrightarrow\infty.
 \label{eq:appC-ML-limit}
\end{equation}

It is convenient to introduce the normalized polynomial
\begin{equation}
 p_L(x)
 =
 \frac{q_L(x)}{M_L}
 =
 \sum_{k=0}^{L-1}
 c_{k,L}\cos\bigl((2k+1)x\bigr),
 \label{eq:appC-pL}
\end{equation}
where
\begin{equation}
 c_{k,L}
 =
 \frac{4}{\pi M_L}
 \frac{(-1)^k}{2k+1}
 \left(
 1-\frac{2k+1}{2L}
 \right).
 \label{eq:appC-coeff}
\end{equation}
By construction,
\begin{equation}
 \|p_L\|_\infty=1.
 \label{eq:appC-pL-norm}
\end{equation}

A second property of central importance for the Bell-CHSH
construction concerns the quadratic weight of the coefficients.
Orthogonality of the cosine functions gives
\begin{equation}
 \frac{1}{2\pi}
 \int_{-\pi}^{\pi}p_L(x)^2\,dx
 =
 \frac{1}{2}
 \sum_{k=0}^{L-1}c_{k,L}^2.
 \label{eq:appC-parseval}
\end{equation}
Since $\|p_L\|_\infty=1$, this immediately implies the Parseval bound
\begin{equation}
 \sum_{k=0}^{L-1}c_{k,L}^2\leq 2.
 \label{eq:appC-parseval-bound}
\end{equation}
For every finite $L$ the inequality is strict. Indeed, equality would
require $p_L(x)^2=1$ almost everywhere. Since $p_L$ is continuous,
this would imply $p_L(x)^2=1$ everywhere, which is impossible for a
nonconstant finite odd-harmonic trigonometric polynomial. Hence
\begin{equation}
 \sum_{k=0}^{L-1}c_{k,L}^2<2
 \qquad
 \text{for every finite }L.
 \label{eq:appC-strict}
\end{equation}

On the other hand, using $M_L\to1$ together with
\begin{equation}
 \sum_{k=0}^{\infty}\frac{1}{(2k+1)^2}
 =
 \frac{\pi^2}{8},
 \label{eq:appC-odd-sum}
\end{equation}
one finds
\begin{align}
 \lim_{L\to\infty}
 \sum_{k=0}^{L-1}c_{k,L}^2
 &=
 \frac{16}{\pi^2}
 \sum_{k=0}^{\infty}
 \frac{1}{(2k+1)^2}
 \nonumber\\
 &=2.
 \label{eq:appC-coeff-limit}
\end{align}
The normalized Fej\'er polynomials therefore provide a sequence of
finite, uniformly bounded, odd-harmonic trigonometric polynomials
satisfying
\begin{equation}
 \|p_L\|_\infty=1,
 \qquad
 \sum_{k=0}^{L-1}c_{k,L}^2<2,
 \qquad
 \sum_{k=0}^{L-1}c_{k,L}^2\longrightarrow2.
 \label{eq:appC-summary}
\end{equation}

These are precisely the properties needed in the Bell-CHSH
construction. The odd-harmonic structure guarantees the exact
anticommutation of the corresponding Weyl-polynomial axis
observables, while the limiting quadratic weight $2$ is responsible
for the approach of the Bell-CHSH correlator to Tsirelson's bound.

\subsection{The Fej\'er normalization}

We now establish explicitly the normalization property
$\|q_L\|_\infty=q_L(0)=M_L$ used above. The Fejér kernel at the order entering \eqref{fejer} is
\begin{equation}
F_{2L-1}(t) = \frac{1}{2L}\left[ \frac{\sin(Lt)}{\sin(t/2)}\right]^2 \;, \nonumber
\end{equation}
so that, in the distributional sense,
\begin{equation}
q_L'(x) = \frac{1}{\pi}\left[ F_{2L-1}(x+\pi/2) - F_{2L-1}(x-\pi/2)\right] \;. \nonumber
\end{equation}
For $0<x<\pi/2$ the squared numerators of the two kernel values coincide, whereas $\sin^2(\pi/4+x/2) > \sin^2(\pi/4-x/2)$, whence $q_L'(x)\le 0$. Since $q_L$ is even, $q_L(x+\pi)=-q_L(x)$ and $q_L(\pi/2)=0$, its uniform norm is $q_L(0)=M_L$, which is the first of eqs.~\eqref{pnorm}.

The Fejér convergence theorem gives $M_L \to q(0)=1$. Dominated convergence, together with $\sum_{m \; {\rm odd}} 1/m^2 = \pi^2/8$, then yields
\begin{equation}
\lim_{L\to\infty} \sum_{k=0}^{L-1} c_{k,L}^{\,2} = \lim_{L\to\infty} \frac{16}{\pi^2 M_L^2} \sum_{k=0}^{L-1} \frac{1}{(2k+1)^2}\left( 1 - \frac{2k+1}{2L}\right)^2 = 2 \;, \nonumber
\end{equation}
which is the third of eqs.\eqref{pnorm}.

\subsection{The six-Weyl polynomial}

With $u=\cos x$, eq.\eqref{psix} reads
\begin{equation}
p_{\rm six}(x) = \frac{13u-16u^3+8u^5}{5} = \frac{u}{5}\left[ 8(u^2-1)^2+5\right] \;. \nonumber
\end{equation}
The polynomial on the right hand side is odd and nonnegative on $[0,1]$, its interior critical points being given by $40u^4-48u^2+13=0$, {\it i.e.}
\begin{equation}
u^2 = \frac{12\pm\sqrt{14}}{20} \;. \nonumber
\end{equation}
At these two points $p_{\rm six} \simeq 0.9969508164$ and $p_{\rm six}\simeq 0.9515273353$, while the endpoint values in the $u$ variable are $0$ and $1$. Oddness then gives $|p_{\rm six}(x)|\le 1$ for $-1\le u \le 1$, {\it i.e.} $||p_{\rm six}||_\infty = 1$.



t
%
%

\section{The degree-$511$ Fej\'er polynomial}
\label{app:p511}

For completeness, we give here the explicit prescription for the
finite Fej\'er polynomial used to obtain the value closest to
Tsirelson's bound reported in the main text.  Taking $L=256$, the
normalized Fej\'er approximant has degree $2L-1=511$ and contains
the $256$ odd harmonics
\begin{equation}
 {\cal M}_{511}=\{1,3,5,\ldots,511\}.
\end{equation}
It can be written as
\begin{equation}
 p_{511}(x)
 =
 \sum_{k=0}^{255}c_{2k+1}^{(511)}
 \cos[(2k+1)x],
 \label{eq:p511}
\end{equation}
with
\begin{equation}
 c_{2k+1}^{(511)}
 =
 \frac{4}{\pi M_{256}}\,
 \frac{(-1)^k}{2k+1}
 \left(1-\frac{2k+1}{512}\right),
 \qquad k=0,\ldots,255,
 \label{eq:p511coeff}
\end{equation}
where
\begin{equation}
 M_{256}
 =
 \frac{4}{\pi}
 \sum_{k=0}^{255}
 \frac{(-1)^k}{2k+1}
 \left(1-\frac{2k+1}{512}\right).
 \label{eq:M256}
\end{equation}
By construction,
\begin{equation}
 \|p_{511}\|_\infty=1 .
\end{equation}

The corresponding observable is therefore a finite Weyl polynomial
containing $512$ Weyl operators, since each cosine is the sum of two
Weyl operators.  No limiting observable is involved at this stage.

In the narrow-band spectral limit, the Bell-CHSH expectation value
associated with a fixed polynomial reduces to
\begin{equation}
 {\cal B}_{511}^{\rm spec}
 =
 \sqrt{2}
 \sum_{k=0}^{255}
 \left(c_{2k+1}^{(511)}\right)^2,
 \label{eq:p511spectral}
\end{equation}
which gives numerically
\begin{equation}
 {\cal B}_{511}^{\rm spec}\simeq 2.80177 .
\end{equation}

The same conclusion is obtained before taking the strict spectral
limit.  For example, choosing
\begin{equation}
 \lambda=0.99999,
 \qquad
 \omega_0=-\frac{\log\lambda}{\pi},
 \qquad
 d=10^{-6},
\end{equation}
and evaluating the exact finite double sum of Eqs.~\eqref{Efin}-\eqref{CHSHfin}, one finds
\begin{equation}
 {\cal B}_{511}\simeq 2.80027 .
 \label{eq:p511finite}
\end{equation}
Thus an entirely finite Weyl polynomial already reaches a Bell-CHSH
value within about one percent of Tsirelson's bound $2\sqrt{2}$.

For the numerical evaluation of Eq.~\eqref{Efin} at such small spectral
parameters it is important to avoid loss of precision from the
cancellation between $\nu_A$, $\nu_B$, and $\kappa$.  We therefore
evaluate the exponent through the exact identity
\begin{equation}
 m^2\nu_A+n^2\nu_B-2mn\kappa
 =
 2\left(
 \frac{m}{\sqrt{u}}-\frac{n}{\sqrt{v}}
 \right)^2+n^2-m^2 ,
 \label{eq:p511stable}
\end{equation}
with $u=1-\mu_+$ and $v=\mu_- -1$.  This form is algebraically
equivalent to the original expression but is numerically stable in
the regime $\lambda\to1$.

\section{Abel-Poisson approximants as an alternative regularization}
\label{app:abel-poisson}

For completeness, we briefly mention another natural approximation
scheme for the square-wave function
\begin{equation}
 q(x)=\operatorname{sgn}(\cos x),
\end{equation}
based on Abel-Poisson summation. This provides an alternative to the
Fej\'er approximants employed in the main text and illustrates that
the odd-harmonic Weyl-polynomial construction is not tied to a unique
summability prescription.

Let
\begin{equation}
 f(x)\sim\sum_{n\in{\mathbb Z}}\widehat f_n e^{inx}
\end{equation}
be the Fourier series of a $2\pi$-periodic integrable function. For
$0\leq r<1$, its Abel-Poisson mean is defined by
\begin{equation}
 {\cal A}_r f(x)
 =
 \sum_{n\in{\mathbb Z}}
 r^{|n|}\widehat f_n e^{inx}.
 \label{eq:AP-definition}
\end{equation}
Equivalently,
\begin{equation}
 {\cal A}_r f(x)
 =
 \frac{1}{2\pi}\int_{-\pi}^{\pi}
 P_r(x-t)f(t)\,dt,
\end{equation}
where
\begin{equation}
 P_r(t)
 =
 \frac{1-r^2}{1-2r\cos t+r^2}
\end{equation}
is the Poisson kernel. Since
\begin{equation}
 P_r(t)>0,
 \qquad
 \frac{1}{2\pi}\int_{-\pi}^{\pi}P_r(t)\,dt=1,
\end{equation}
Abel-Poisson summation is contractive,
\begin{equation}
 \|{\cal A}_r f\|_\infty\leq\|f\|_\infty,
\end{equation}
and ${\cal A}_r f(x)\to f(x)$ as $r\uparrow1$ at every continuity
point of $f$.

For
\begin{equation}
 q(x)=\operatorname{sgn}(\cos x)
 =
 \frac{4}{\pi}
 \sum_{k=0}^{\infty}
 \frac{(-1)^k}{2k+1}
 \cos\bigl((2k+1)x\bigr),
\end{equation}
the Abel-Poisson approximant takes the particularly simple form
\begin{equation}
 q_r(x)
 =
 \frac{4}{\pi}
 \sum_{k=0}^{\infty}
 \frac{(-1)^k r^{2k+1}}{2k+1}
 \cos\bigl((2k+1)x\bigr).
 \label{eq:AP-square-wave}
\end{equation}
Thus Abel-Poisson summation preserves exactly the odd-harmonic
structure which is responsible for the anticommutation property used
throughout the present construction. Moreover,
\begin{equation}
 q_r(x)
 =
 \frac{2}{\pi}
 \arctan\left(
 \frac{2r\cos x}{1-r^2}
 \right),
\end{equation}
so that
\begin{equation}
 q_r(x)\longrightarrow\operatorname{sgn}(\cos x),
 \qquad r\uparrow1,
\end{equation}
away from the discontinuity points.

For applications involving finite Weyl polynomials one may truncate
(\ref{eq:AP-square-wave}) after $L$ odd harmonics,
\begin{equation}
 q_{r,L}(x)
 =
 \frac{4}{\pi}
 \sum_{k=0}^{L-1}
 \frac{(-1)^k r^{2k+1}}{2k+1}
 \cos\bigl((2k+1)x\bigr).
 \label{eq:AP-finite}
\end{equation}
For instance, $L=6$ gives a polynomial of degree $11$, involving the
harmonics $1,3,5,7,9,11$, and therefore the same number of Weyl
operators as the degree-$11$ Fej\'er polynomial. The corresponding coefficients are exponentially damped
as $r^m$, rather than by the linear Fej\'er weights.

A minor but important distinction should be noted. While the complete
Abel-Poisson mean satisfies $\|q_r\|_\infty\leq1$, a finite
truncation $q_{r,L}$ does not automatically inherit this bound.
Accordingly, when it is used to define a bounded observable, one may
introduce the normalized polynomial
\begin{equation}
 p^{\rm AP}_{r,L}(x)
 =
 \frac{q_{r,L}(x)}
      {\|q_{r,L}\|_\infty},
 \qquad
 \|p^{\rm AP}_{r,L}\|_\infty=1.
 \label{eq:AP-normalized}
\end{equation}

The Abel-Poisson family therefore provides another natural class of
finite odd-harmonic Weyl polynomials. Although the present work uses
Fej\'er approximants because their finite-degree norm properties are
particularly convenient, Eq.~(\ref{eq:AP-finite}) shows that the same
algebraic mechanism can be explored with alternative harmonic
regularizations. A systematic comparison of such families, including
their efficiency at fixed polynomial degree in approaching maximal
Bell-CHSH violation, may be of independent interest.

\end{document}